# Frequency Control in Microgrids: An Adaptive Fuzzy-Neural-Network Virtual Synchronous Generator

Waleed Breesam, Rezvan Alamian, Nima Tashakor, Brahim Elkhalil Youcefa, and Stefan M. Goetz

*Abstract*—The reliance on distributed renewable energy has increased recently. As a result, power electronic-based distributed generators replaced synchronous generators which led to a change in the dynamic characteristics of the microgrid. Most critically, they reduced system inertia and damping. Virtual synchronous generators emulated in power electronics, which mimic the dynamic behaviour of synchronous generators, are meant to fix this problem. However, fixed virtual synchronous generator parameters cannot guarantee a frequency regulation within the acceptable tolerance range. Conversely, a dynamic adjustment of these virtual parameters promises robust solution with stable frequency. This paper proposes a method to adapt the inertia, damping, and droop parameters dynamically through a fuzzy neural network controller. This controller trains itself online to choose appropriate values for these virtual parameters. The proposed method can be applied to a typical AC microgrid by considering the penetration and impact of renewable energy sources. We study the system in a MATLAB/Simulink model and validate it experimentally in real time using hardware-in-the-loop based on an embedded ARM system (SAM3X8E, Cortex-M3). Compared to traditional and fuzzy logic controller methods, the results demonstrate that the proposed method significantly reduces the frequency deviation to less than 0.03 Hz and shortens the stabilizing/recovery time.

## Nomenclature

### A. Parameters

| | |
|---|---|
| $\Delta P_m$ | Generated power from the turbine system |
| $\Delta P_g$ | Change signal of governor control action |
| $\Delta P_c$ | Action change signal from secondary control |
| $R$ | Governor droop constant |
| $\Delta f$ | Frequency deviation |
| $K$ | Gain of the integral controller |
| $\beta$ | Bias factor |
| $\Delta P_w$ | Generated power change from the wind turbine system |
| $\Delta P_{\text{wind}}$ | Initial wind power change |
| $\Delta P_{pv}$ | Generated power change from the photovoltaic panel system |
| $\Delta P_{\text{solar}}$ | Initial solar power change |
| $\Delta P_L$ | Electrical load power change |
| $T_t$ | Turbine time constant |
| $T_g$ | Governor time constant |
| $T_w$ | Wind turbine time constant |
| $T_{pv}$ | Solar system time constant |
| $\Delta P_{vi}$ | Output power change from an energy storage system-based inverter |
| $k_v$ | Virtual inertia characteristic constant |
| $D_v$ | Virtual damping constant |
| $T_{\text{ESS}}$ | Time constant of inverter-based energy storage |
| $R_v$ | Virtual droop constant |
| $E$ | Error |
| $y_d$ | Desired response |
| $y$ | Actual output |
| $e$ | Error signal |
| $w_{ijk}$ | Connection weights between layers |
| $m_{ijk}$ | Center with respect to $A_{ij}$ membership function |
| $\sigma_{ijk}$ | Width with respect to $A_{ij}$ membership function |
| $\eta_w, \eta_m, \eta_\sigma$ | Learning-rates of the adjusted parameters of FNNC |
| $\delta_{j4}, \delta_{j3}$ | Error propagating in layers |
| $O_{ij2}$ | Output from memberships |

### B. Abbreviations

| | |
|---|---|
| VSG | Virtual Synchronous Generator |
| FNNC | Fuzzy Neural Network Controller |
| RoCoF | Rate of Change of Frequency |
| PV | Photovoltaic |
| ESS | Energy Storage System |
| GRC | Generation Rate Constraint |

## I. Introduction

Renewable energy based distributed generation such as photovoltaics and wind turbines promises to solve the environmental problems of fossil energy [1], [2]. In contrast to conventional fossil power, however, renewable energy is almost exclusively converted and injected into a microgrid through power electronics, which lack rotational inertia [3], [4]. This rotational inertia served as an ultra-short-term storage and stabilized both frequency and voltage in previous power systems. A further growth of inverter-based renewable energy can destabilize microgrids and lead to frequency fluctuations.

The absence of physical inertia is one of the causes of increased fluctuations in the system. However, inverters could be controlled to mimic the dynamic behaviour of synchronous generators and provide synthetic constant and damping coefficient through so-called virtual synchronous generators (VSG) [5, 6, 7].

Several studies have comprehensively compared VSG topology [8, 9]. One of the most common VSG techniques for synchronous generator emulation is the swing equation-based method [10, 11]. This equation relies on inertia and damping parameters, which can be combined with droop coefficient characteristics. Droop plays a key role in control algorithms by enabling power sharing between inverters and improving frequency responses if its value is accurately determined. Selecting appropriate VSG parameters in controller design is essential because of their mutual influence on each other as well as the frequency response and stability of the microgrid [12, 13].

Some recent studies have presented frequency regulation approaches based on fixed VSG parameters on the systems which involve inverter-interfaced distributed generators [14, 15, 16, 17]. However, these studies do not explain or clarify the performance and effectiveness of VSG when its parameters are changed. Moreover, adopting fixed values for VSG parameters does not reflect the actual or necessary feasibility of selecting optimal parameter values under various microgrid conditions.

On the other hand, Xu et al. offer an alternative approach to address the optimal selection of VSG parameters through an improved algorithm [18]. This algorithm modifies the effect of the damping factor by cascading a differential link into different positions of a first-order virtual-inertia forwarding channel. However, this algorithm relies on a mathematical analysis instead of actual considerations to select VSG parameters. In addition, it does not account for the effects of disturbances and uncertainties that may occur in the system over time.

It is worth to mentioning that fixed VSG parameters do not guarantee the regulation of the microgrid frequency within a tight range, especially in the presence of disturbances and uncertainties. To strictly regulate microgrid frequency, the VSG parameters should be dynamically adjusted [19, 20]. To this end, several studies have introduced advanced techniques. For instance, model-predictive control can adjust virtual inertia and damping, in which the output of the controller is represented by tuned virtual parameters and the input by the frequency deviation [21]. Despite its effectiveness, this method relies on trial and error to select the parameters. It also requires discrete time representation of the microgrid model to predict future states. This limitation leads to the requirement of prior knowledge of the system, along with the same problem of optimal initial value selection in controller design.

In response to this limitation, a study developed a data-driven approach with reinforcement learning to adjust virtual inertia and damping without prior knowledge of the system [22]. This method requires the definition of priorities in the reward function before the controller design process, e.g., frequency deviation and rate of change of frequency (RoCoF) prioritized at the expense of frequency settling time. This method leads to limitations in the flexibility of control actions, especially in unexpected and sudden events.

Mohammed et al. proposed a method for adapting VSG parameters through real-time estimation of grid impedance [23]. This approach relies on a mathematical analysis of a VSG-based grid forming inverter and the estimation of grid impedance to determine the values of virtual parameters. This design requires prior knowledge of the system model and the measurement of the grid impedance, which adds complexity to the proposed method. On the other hand, a cooperative adaptive control involving inertia, damping, and droop coefficients was presented to establish a small-signal model for dual-parallel VSGs [24]. This method requires measuring the output impedance of VSG and load. Measuring system loads is a challenging process because they are often regarded as unmeasurable disturbances, which further complicates the suggested method. Moreover, the process of adapting virtual parameters requires the addition of a new parameter specific to the proposed adaptation law, which in turn necessitates precise specification. Again, this limitation leads to the same problem of optimal initial value selection in the controller design stage.

Another solution is to adapt the virtual parameters of VSG by using artificial intelligence to maintain frequency regulation and microgrid stability. These advanced methods have received great attention recently, outperforming many other similar methods due to adaptation capability, prediction, and modelling of complex system behaviour without requiring prior knowledge [25, 26]. Such methods can solve nonlinear problems and uncertainties as present in microgrid systems.

A fuzzy logic controller can adjust virtual inertia based on frequency deviation [12] or can be used to augment a governor's output power through an additional term to increase the system's inertia during transients [27]. However, these studies do not account for the other influencing parameters: the selection of optimal damping and droop parameters under different conditions.

In the same context, a study introduces an enhanced approach based on the fuzzy logic controller for managing frequency regulation in an isolated microgrid. This approach uses the frequency deviation and the power of the distributed generator as input to adapt the inertia and damping parameters of the VSG-based inverter dynamically at the same time with virtual parameters of the primary and secondary control loop of the power system [28]. Despite its effectiveness, however, it requires prior knowledge of the system's behaviour to design fuzzy control rules, especially since things become more complicated if the number of parameters to be adapted increases. Furthermore, it is not always robust enough when compared to other artificial intelligence methods such as neural network.

Yao et al. used a radial-basis neural network to adaptively adjust inertia and damping [29]. The authors used the frequency and its derivative as inputs for the neural network, with only virtual inertia as the output. In addition, they modified damping depending on the value of inertia. The method assumes that the system behavior is known in mathematical form and is therefore regarded as a nonintegral adaptation process. This means that the damping parameter values are not flexible, since the parameter is not independently adapted and is rather constrained by the adaptation of the inertia parameter. In response to this gap, Ling et al. proposed an input–output radial basis function neural network to adaptively adjust virtual inertia and damping [3]. This network relies on offline training, which means that it requires a system-related data set to train the neural network before its operation. This technique is rather limited, especially because it involves a specific data set that may not include data on new or sudden changes. In addition, its effectiveness has not yet been tested under conditions of load change.

To overcome the above challenges, we propose a novel technique that combines the capabilities of fuzzy logic control and artificial neural networks to adjust the three virtual parameters: inertia, damping, and droop. The fuzzy neural network controller (FNNC) is capable of functioning with complex and nonlinear systems that require prediction and adaptation under different operating conditions, especially during online training where the most important feature provided by online learning is the ability to deal with the dynamic changes of microgrid systems. In contrast to offline learning, which requires prior knowledge of the system's behavior through the dataset, this controller does not rely on mathematical analysis and system models to determine the values of the virtual parameters. In addition, it is capable of adapting to changing parameters and uncertainties that may occur in the system, such as renewable-energy-dominated microgrids, which typically experience environmental changes. Given their dynamic adaptability, FNNC is regarded as a feasible technique to minimize frequency deviations and shorten the recovery time in microgrids. The main contributions in this paper are as follows:

1- We propose an FNNC with online training capability to simultaneously and dynamically adjust the virtual parameters of a VSG-based inverter. This approach introduces a next generation of VSG-based inverter with flexible, predictive, and adaptive, features. It provides synthetic adaptive characteristics of synchronous generators for the microgrid under disturbances and transient states. Also, it enables a substantial reduction the frequency deviation to less than 0.03 Hz and shortens the stabilizing/recovery time.

2- We present a new type of model-free virtual control based on fuzzy neural network. This technique does not require prior knowledge of the dynamic behavior of the microgrid to precisely regulate its frequency. Instead, it relies on minimal measurements (frequency only) to independently complete its tasks. In addition, it does not require further calculations or measurements, indicating its simple design and high performance. We experimentally test and validate the control algorithm in real time on an embedded ARM system (SAM3X8E, Cortex-M3).

The rest of this paper is structured as follows: Section II presents a small-signal model of an islanded microgrid with conventional virtual inertia. Section III discusses the application of adaptive parameters using a fuzzy controller. Section IV describes the proposed method based on FNNC in detail. Section V presents the results by comparing the above methods. Finally, Section VI concludes the paper.

## II. MODELING OF MICROGRIDS

Microgrids include various distributed generators and renewable energy sources. However, because of their distributed and intermittent nature, modeling their behavior is a challenging process. At a low degree of disturbance, the dynamic equations describing the power system response can be linearized for analysis [30, 31]. Fig. 1 illustrates a typical small-scale microgrid power system. This system consists of a 20-MW synchronous generator, a 6-MW photovoltaic source, an 8-MW wind turbine, a 20-MW industrial load, and a 4-MW storage-based inverter equipped with a virtual synchronous generator control unit.

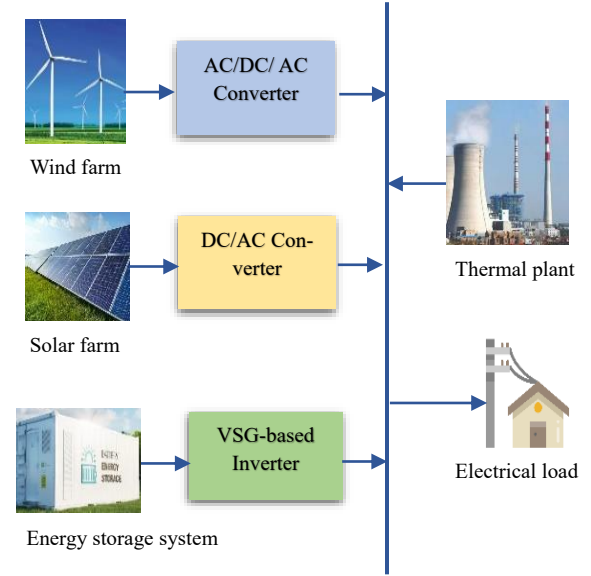

Fig. 1. A typical small-scale microgrid.

The block diagram of the considered system is presented in Fig. 2. The dynamic small-signal model of the microgrid system in the frequency domain $S$ is

$$\Delta P_m(s) = \frac{\Delta P_g(s)}{1 + sT_t}, \quad (1)$$

$$\Delta P_g(s) = \frac{\Delta P_c(s) - \frac{1}{R}\Delta f(s)}{1 + sT_g}, \quad (2)$$

$$\Delta P_c(s) = \frac{K}{s} \beta \cdot \Delta f(s), \quad (3)$$

$$\Delta P_w(s) = \frac{\Delta P_{\text{wind}}(s)}{1 + sT_w}, \quad (4)$$

$$\Delta P_{pv}(s) = \frac{\Delta P_{\text{solar}}(s)}{1 + sT_{pv}}, \quad (5)$$

$$\Delta f(s) = \frac{\Delta P_m(s) + \Delta P_w(s) + \Delta P_{pv}(s) + \Delta P_{vi}(s) - \Delta P_L(s)}{2Hs + D}, \quad (6)$$

where $\Delta P_m$ is the generated power from the turbine system, $\Delta P_g$ the change signal of governor control action, $\Delta P_c$ the action change signal from secondary control, $R$ governor droop constant, $\Delta f$ the frequency deviation, $K$ the gain of the integral controller, $\beta$ the bias factor, $\Delta P_w$ the generated power change from the wind turbine system, $\Delta P_{wind}$ the initial wind power change, $\Delta P_{pv}$ the generated power change from the photovoltaic panel system, $\Delta P_{solar}$ the initial solar power change, and $\Delta P_L$ the electrical load power change. $T_t$, $T_g$, $T_w$, and $T_{pv}$ are respectively the time constants of turbine, governor, wind, and solar systems. $\Delta P_{vi}$ is the output power change from an energy storage system-based inverter [32].

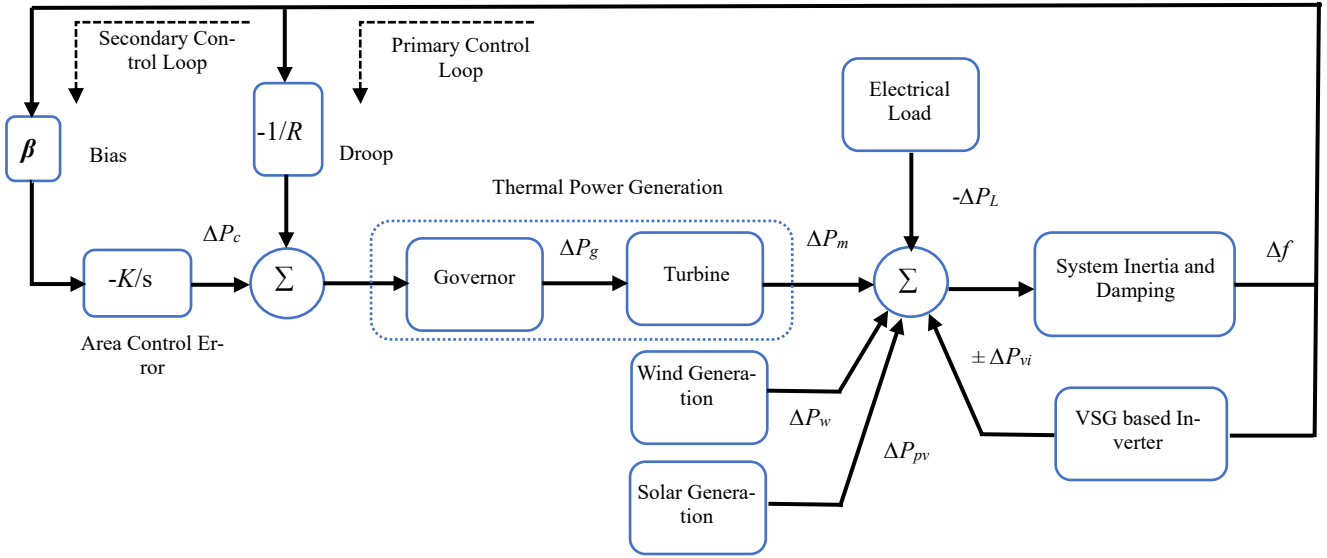

Fig. 2. Block diagram of small-signal model of microgrid system.

The virtual control unit synthetically mimics the dynamic characteristics of synchronous machines to compensate for low inertia and damping resulting from inverter-based renewable energy sources. This unit controls an inverter to inject power into the system or absorb power from it to increase overall inertia and damping as well as improve frequency response and stability of the microgrid

$$\Delta p_{vi}(s) = \frac{(k_v S + D_v)\,\Delta f(s)}{1 + sT_{ESS}}, \quad (7)$$

where $k_v$ is the virtual inertia characteristic constant, $D_v$ the virtual damping constant, and $T_{ESS}$ the time constant of inverter-based energy storage.

If droop is considered in the control unit, then the above equation transforms into

$$\Delta p_{vi}(s) = \frac{(k_v S + D_v)}{1 + sT_{ESS}} \cdot \frac{\Delta f(s)}{R_v}, \quad (8)$$

where $R_v$ is the virtual droop constant that controls active power with frequency regulation.

Setting the values of inertia, damping, and droop in the control unit is important to evaluate the system's performance, response, and stability. The proposed method updates these values adaptively.

## III. ADAPTIVE VIRTUAL PARAMETERS USING FUZZY LOGIC CONTROLLER

In a fuzzy controller, expert data are modeled using linguistic IF-THEN rules, which in turn establish a fuzzy logic inference system. Learning is demonstrated by the establishment of rules under the supervision of experts [33]. Given that microgrid systems are typically dominated by inverter-based sources (wind and solar power), changes in grid loads and renewable energy sources are regarded as disturbances. In this study, we only measured the changes observed in renewable energy power and regarded load changes as unmeasurable disturbances. Fig. 3 shows the structure of the fuzzy controller -based adaptive VSG parameters.

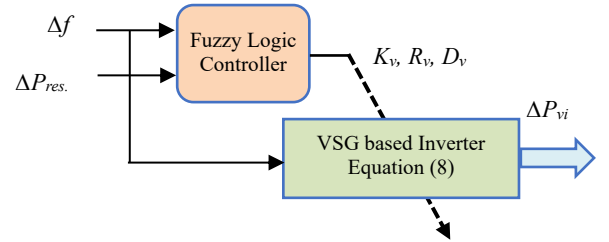

Fig.3. The structure of fuzzy controller-based adaptive VSG parameters.

The fuzzy controller has two inputs: frequency deviation, and the total power changes in renewable energy sources. The output of the controller includes three values, which represent the virtual parameters that require simultaneous adjustment. The input range of the fuzzy variables is $\Delta f = [-0.5, 0.5]$ Hz, with $\Delta P_{res.} = [0, 0.1]$ per unit, where $\Delta P_{res.}$ is the total change in power generated from renewable energy sources.

To design fuzzy control rules, it is important to understand the relationship between the three parameters of inertia, damping, and droop and their relationship with frequency deviation and the total change in power generated from renewable energy sources. Changes in the damping coefficients should be directly proportional to changes in inertia. These changes reduce the settling time, but they generate overshoot in response. Conversely, reducing the droop coefficient mitigates frequency deviations but increases the settling time. These relationships can be used to design fuzzy rules. Tables I–III present the fuzzy rules established for the linguistic variables of the proposed multi-parameter adjustment.

Table I
Fuzzy rules of virtual inertia control for linguistic variables

|  |  | $\Delta f$ | | | | |
|---|---|---|---|---|---|---|
|  |  | VN | N | Z | P | VP |
|  | L | VS | VS | VS | VS | M |
| $\Delta P_{res.}$ | M | S | S | M | H | H |
|  | H | M | M | M | VH | VH |

Table II
Fuzzy rules of virtual droop coefficient control for linguistic variables

|  |  | $\Delta f$ | | | | |
|---|---|---|---|---|---|---|
|  |  | VN | N | Z | P | VP |
|  | L | VS | S | M | S | VS |
| $\Delta P_{res.}$ | M | VS | S | M | S | VS |
|  | H | VS | S | M | S | VS |

Table III
Fuzzy rules of virtual damping control for linguistic variables

|  |  | $\Delta f$ | | | | |
|---|---|---|---|---|---|---|
|  |  | VN | N | Z | P | VP |
|  | L | S | S | S | S | M |
| $\Delta P_{res.}$ | M | S | S | M | H | H |
|  | H | M | M | M | H | H |

where N is negative, VN is very negative, Z is zero, P is positive, VP is very positive L is low, S is small, VS is very small, M is medium, H is high, VH is very high.

For example, if $\Delta f$ is VN and the $\Delta P_{res.}$ supplying the microgrid is L, the values of $k_v$ and $D_v$ will be VS and S respectively. Conversely, if $\Delta f$ is VP and $\Delta P_{res.}$ supplying the microgrid is H, this means that the values of $k_v$ and $D_v$ must be VH and H. For the virtual parameter $R_v$, its value should be reduced as $\Delta f$ value increases positively or negatively.

It is preferable to choose the ranges of the VSG parameters in a reasonable and appropriate manner compared to the case when fixed parameters are used, to support the system with flexible ability and more synthetic dynamic characteristics under different operating conditions. In other words, the selection of upper limit values for $k_v$ and $D_v$ should be high enough, while the lower limit is expected to be not very low. In contrast, the minimum limit of $R_v$ is much lower to provide a less frequency deviation. Based on this, the range of $k_v$ and $D_v$ and $R_v$ are set by trial and error, taking into account the balance between the above requirements as follows: [0.5, 7], [0.1, 10], and [0.005, 2.7] respectively.

## IV. PROPOSED METHOD BASED ON FNNC

### A. Basics of FNNC

Modern controllers, such as robust adaptive controllers with a variable structure, rely on detailed mathematical models, which is not always possible in a bias-free was for time-varying and highly nonlinear plants [34, 35]. We use an FNNC, which combines the advantages of both neural networks (e.g., optimization abilities, learning capabilities, and connectionist structures) and fuzzy systems (e.g., human-like IF-THEN rules and expert knowledge). Figure 4 illustrates the basic structure of a four-layer FNN. The basic function of each layer is described in the following.

1- Input Layer
This layer transmits the input signals ($x_i$) to the output without any change.

2- Membership layer
In this layer, each node performs a gaussian membership function:

$$o_{ij}^2 = exp\left(-\frac{(x_j - m_{ij})^2}{\sigma_{ij}^2}\right) \quad (9)$$

where $m_{ij}$ and $\sigma_{ij}$ denote the center and width with respect to $A_i^j$ membership function.

3- Rule layer
The output of each node in this layer is the product of the incoming signals from second layer and is denoted by $\Pi$.

$$o_j^3 = \prod_{i=1}^{n} x_{ij}^3 \quad (10)$$

4- Output Layer
Each node in this layer output the summation of all incoming signals

$$o_j^4 = \sum_{i=1}^{m} w_{ij}^4 \cdot x_i^4 \quad (11)$$

We defined fitness function to describe the online learning algorithm of this FNNC on the basis of back-propagation as

$$E = \frac{1}{2}(y_d - y)^2 = \frac{1}{2}e^2, \quad (12)$$

where $y_d$ is the desired response, $y$ the actual output, and $e$ the error signal.

To minimize the error between the actual system output and the reference output in each iteration, the learning algorithm adjusts the connection weights between Layers 3 and 4 and the parameters of the membership functions of the FNNC (center and width) in Layers 3. The weights and parameters of the membership functions are updated according to

$$w_{ij}(k+1) = w_{ij}(k) + \eta_w \delta_j^4 x_i^4, \quad (13)$$

$$m_{ij}(k+1) = m_{ij}(k) + \eta_m \delta_j^3 \left(\prod_{i=1}^{n} O_{ij}^2\right) \frac{2(x_{ij}^2 - m_{ij})}{\sigma_{ij}^2}, \quad (14)$$

$$\sigma_{ij}(k+1) = \sigma_{ij}(k) + \eta_\sigma \delta_j^3 \left(\prod_{i=1}^{n} O_{ij}^2\right) \frac{2(x_{ij}^2 - m_{ij})^2}{\sigma_{ij}^3}, \quad (15)$$

$$\delta_j^3 = \sum_{j=1}^{p} \delta_j^4 w_{ij}^4, \qquad (16)$$

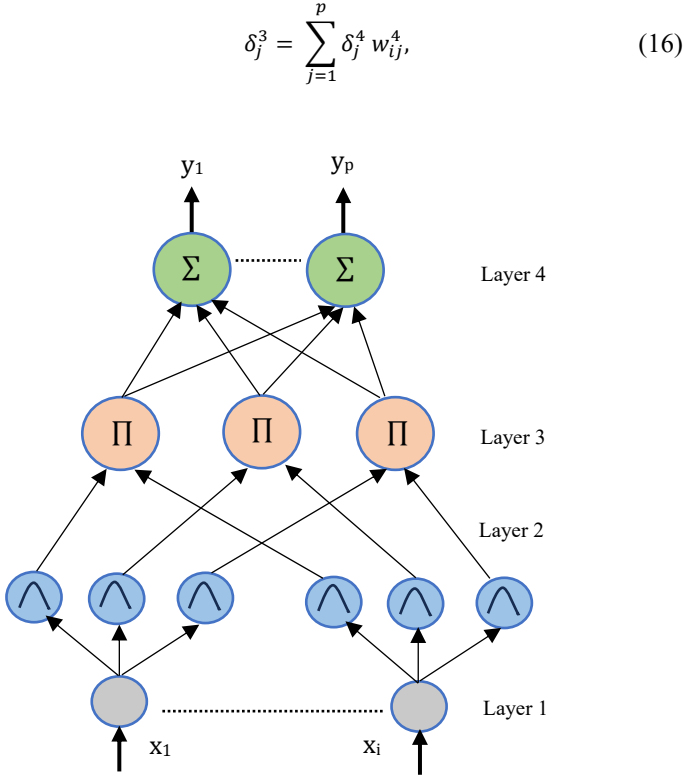

Fig. 4. The structure of four-layer FNN.

where $\delta_j^4$ is error propagating in Layer 4, $x_i^4$ are the input signals to Layer 4, $O_{ij}^2$ the output from memberships, $\delta_j^3$ is error propagating in Layer 3, $m_{ij}$ is center with respect to $A_i^j$ membership function, $\sigma_{ij}$ is width with respect to $A_i^j$ membership function, $\eta_w$, $\eta_m$, and $\eta_\sigma$ are the learning-rate of the adjusted parameters of the FNNC, $p$ is the number of output nodes, and $k$ is current iteration step. The expression of $\delta_j^4$ follows

$$\delta_j^4 = Ae + \frac{de}{dt}, \qquad (17)$$

where A is a positive constant.

It is worth noting that determining the values of the learning rates is significant for training the FNNC effectively. If small values are chosen, convergence is guaranteed, but at a slow speed. On the other hand, if large values are chosen, instability may occur [36]. In fact, the convergence and stability of the FNNC is proofed in details based on the analysis of Lyapunov stability, and adaptive learning rates is presented [37]. For simplicity, however, small fixed values will be adopted for the learning rates in this paper.

### B. FNN implementation for virtual parameters control

The proposed FNNC adjust the three virtual parameters online through back-propagation algorithm (Fig. 5). The frequency deviation forms the first input, RoCoF the second. Each input of the FNNC had three Gaussian membership functions. Adjusting was conducted across the three virtual parameters until the error reached an acceptable degree of tolerance, i.e.,
about $10^{-3}$.

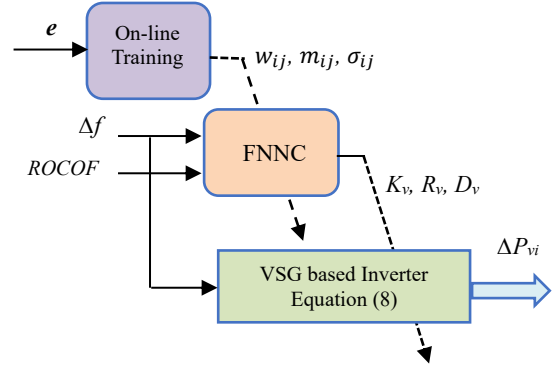

Fig. 5. Proposed structure of FNNC based-adaptive VSG parameters.

### V. RESULTS AND DISCUSSION

We modelled the microgrid in MATLAB/Simulink on the host PC (Fig. 2). Table IV presents the model parameters. Several scenarios study the proposed method. In these scenarios, changes in load, variations in the power of the renewable energy source, and internal variations in the system characteristics (uncertainties) form the key disruptions. Changes in wind turbine capacity and solar energy form measurable disturbances. We validated the results using a hardware-in-the loop based on an embedded system (SAM3X8E Cortex M3). We implemented the control algorithm (fuzzy controller and FNNC) on the microcontroller in C++ and allowed it to interact with the microgrid model in real-time through an input-output interface unit. The microcontroller signal, waveform, and responses are monitored and recorded by the host PC in real-time. Fig. 6 represents a block diagram of the proposed system's configuration. The learning rates of the adjusted FNNC parameters were 0.2, 0.02, and 0.01. Fig. 7 shows the flowchart for the FNNC implementation.

Table IV
Microgrid system Parameters for Scenario I and II.

| Parameter | Symbol | Unit | Value |
| --- | --- | --- | --- |
| Gain of Integral controller | $K$ | s | 0.2 |
| Governor time constant | $T_g$ | s | 0.1 |
| Turbine time constant | $T_t$ | s | 0.4 |
| Governor droop constant | $R$ | Hz/p.u.MW | 2.4 |
| Bias factor | $B$ | p.u.MW/Hz | 0.99 |
| System inertia constant | $H$ | p.u.MWs | 0.082 |
| System load damping | $D$ | p.u.MW/Hz | 0.016 |
| Time constant of wind turbine | $T_w$ | s | 1.4 |
| Time constant of inverter-based ESS | $T_{ESS}$ | s | 5 |
| Time constant of solar system | $T_{pv}$ | s | 1.9 |
| Fixed virtual inertia constant | $K_v$ | p.u. s | 1.3 |
| Fixed virtual damping constant | $D_v$ | p.u.MW/Hz | 1.2 |
| Fixed virtual droop constant | $R_v$ | Hz/p.u.MW | 2.7 |
| maximum capacity of energy storage system-based inverter | | p.u. | ± 0.29 |

***Scenario I*: Basic Assessment of Fuzzy Controller and FNNC**

The first scenario connects the distributed generators (wind turbine and solar energy) to the microgrid of +0.1 p.u. at $t = 5$ s, they were disconnected at $t = 60$ s. Load step changes of +0.1 p.u. occur at $t = 20$ s and –0.1 p.u. occur at $t = 40$ s. Fig.

7 compares the frequency deviations based on the fuzzy controller for three methods: adaptive inertia only, inertia and damping only, and adapting all parameters (inertia, damping, and droop.

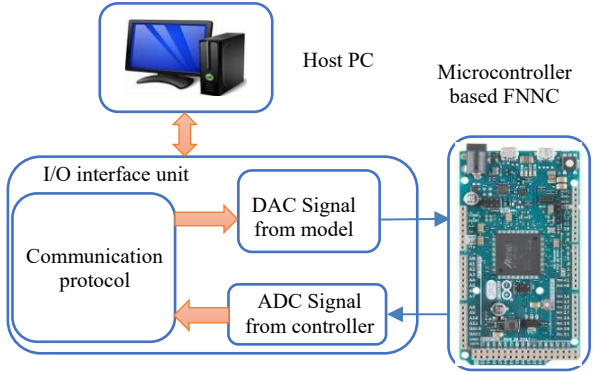

Fig. 6. Block diagram of the system configuration.

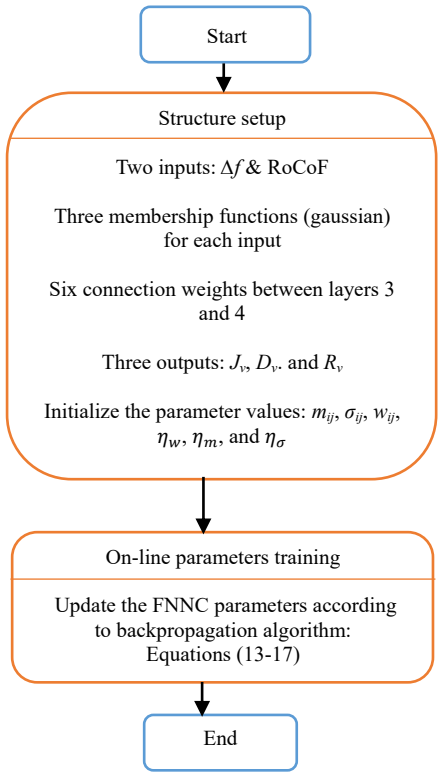

Fig. 7. Flowchart of FNNC implementation.

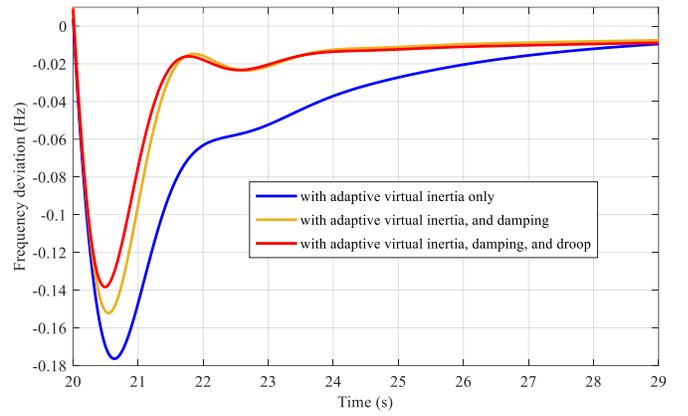

Fig. 7. Response of the frequency deviation.

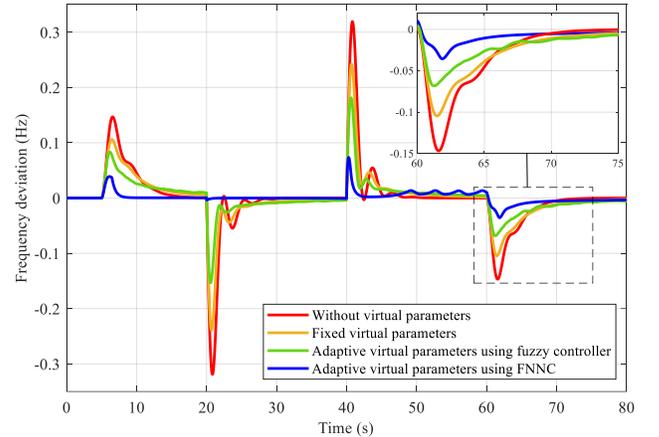

Fig. 8. Response of the frequency deviation for step changes in load and renewable energy sources.

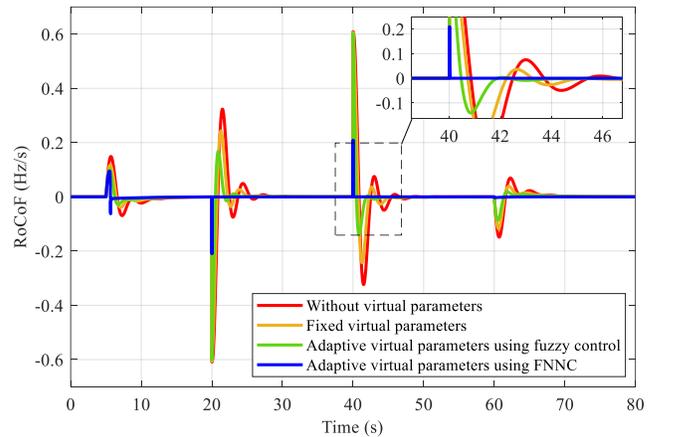

Fig. 9. RoCoF response for step changes in load and renewable energy sources.

It can be noted that dynamically adjusting the three parameters (inertia, damping and droop) improves the frequency response significantly compared to adapting the virtual inertia only, and slightly compared to adapting both inertia and damping parameters. Figure 8 in turn graphs the frequency responses of the aforementioned disturbances while Fig. 9 shows the RoCoF for Scenario I. It is clear that the fuzzy controller significantly reduced the frequency deviation to less than half compared to conventional methods, whereas the FNNC faired best and reduced the frequency deviation by another factor of ½. The online training process took about 2 s for the step change at at $t = 5$ s, while it took about 7 s when dealing with the step change at at $t = 60$ s. On the other hand, RoCoF within a range of 0.5 Hz/s to 1 Hz/s are generally adopted in power systems. This means that the proposed method reduces RoCoF effectively compared to previous methods, which have a risk if the power system is based on a value of 0.5 Hz/s, as RoCoF exceeding the permissible limit will trigger the relays and lead to sequential trips in the system. The VSG-based inverter played a key role in the injection or absorption of power into or from the microgrid (Fig. 10). During the period from 5 to 20 s, for instance, the inverter absorbed power from the microgrid to compensate for the increase in power resulting from the renewable energy sources. Conversely, during the period from 20 to 40 s, the inverter supplied power to the microgrid to compensate for the increased load. This mechanism supports the microgrid system and mitigates the strain. Also, it improves the system's response by reducing its frequency deviation and stabilizing time.

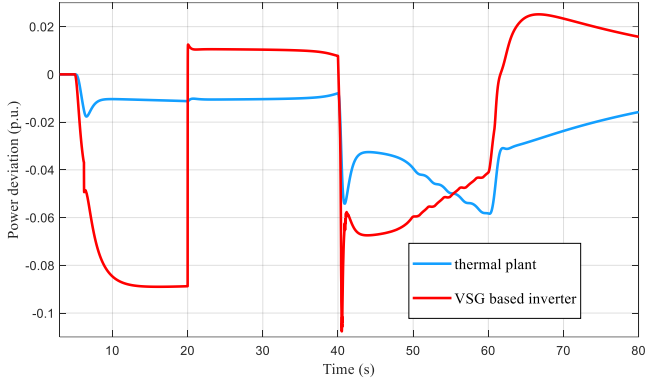

Fig. 10. Active power of thermal plant and VSG based inverter.

*Scenario II*: **Performance Assessment of the Proposed FNNC Under Low and High Penetration of Renewable Energy Sources**

Scenario II used wind and solar power to test the system's frequency response in two cases: low renewable energy penetration (with high load) and high renewable energy penetration (with low load). In the first case, the ranges of wind and solar powers are respectively 0.1–0.12 p.u. and 0.08–0.09 p.u. In the second case, the ranges of wind and solar power are respectively 0.2–0.3 p.u. and 0.1–0.2 p.u. These ranges of renewable energy source capacity account for power variations over time with changing environmental conditions.

Figure 11 illustrates the first case of the fuzzy logic controller and FNNC at a load varying between 0.1 and 0.3 p.u. FNNC keeps the range of frequency deviation within ±0.02 Hz, whereas the fuzzy controller requires more than twice as much and causes ±0.05-Hz swings. As a result of the system's dynamically adaptive parameters, which can accommodate changes and disturbances occurring over time, the system was not greatly affected by the instantaneous changes that occurred in the microgrid.

The second case tested continuous load changes within 0.05–0.1 p.u. and connected wind as well as solar power to the microgrid at *t* = 10 s (Fig. 12). At *t* = 50 s, solar power was disconnected from the microgrid. Again, the FNNC outperformed the fuzzy controller by more than a factor of two. Specifically, it successfully learned the dynamic characteristics of the microgrid system during online training.

Figure 13 shows RoCoF response for Scenario II (second case), where RoCoF of the proposed method remained within the range ± 0.2 Hz/s. Also, Fig. 14 illustrations how the inverter based on FNNC deals with the fluctuations and dominance of renewable sources characteristics on the microgrid by absorbing or injecting the largest possible power compared to other methods.

*Scenario III*: **System Performance Under Physical Constraints, Communication Time Delay, and Uncertainties**

The generation rate constraint and the speed-governor dead band represent the most critical physical constraints in conventional electric power plants. Time delays typically occur during the frequency filtering process and measurement in the second control loop.

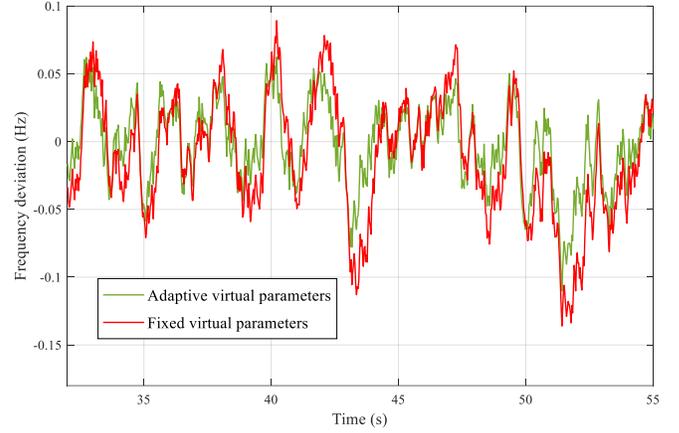

(a)

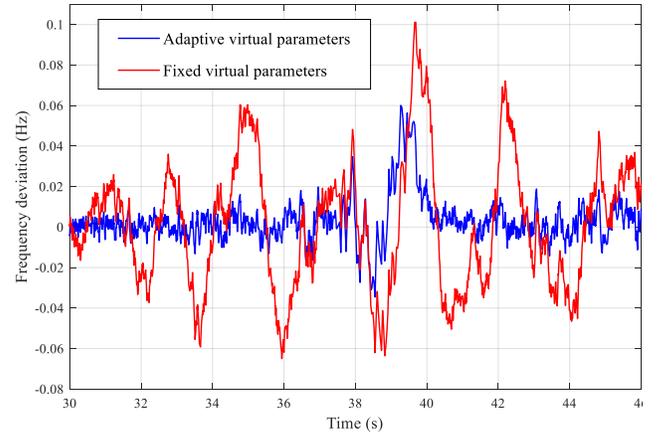

(b)

Fig. 11. The response of frequency deviation low integration of RES and high load changes using: (a) using fuzzy controller (b) using FNNC.

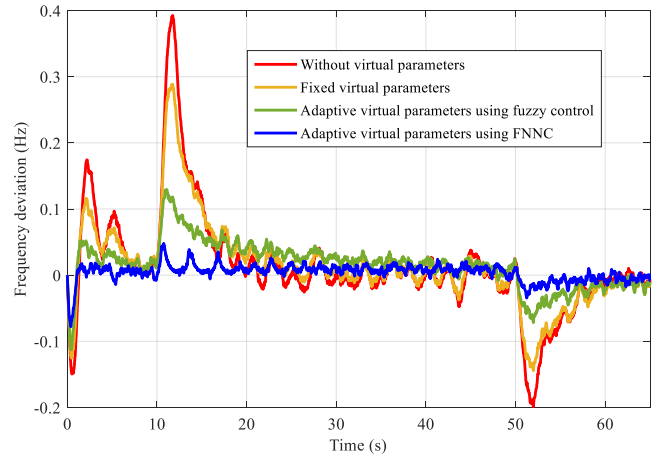

Fig. 12. The response of frequency deviation for high integration of renewable energy source and low load changes.

Uncertainties also play a key role in this process, during which the values of basic microgrid parameters, namely system inertia, system damping, and governor and turbine time constants, change. Scenario III examined these factors to determine their effect on the performance of the proposed control method. Table V presents the system parameters used in this scenario. Fig. 15 depicts changes in frequency deviation response in the presence of physical constraints, communication time delays for secondary control, and uncertainties. Scenario III uses the same temporal load and energy source profiles as the basic test in Scenario I. Again, the proposed control using FNNC achieved high performance in reducing frequency deviations and improving system response with lower stabilizing time. It reduced the frequency deviation to less than $1/2$ compared to an adaptive fuzzy control and almost $1/4$ compared to constant parameters. It was also not affected by uncertainties or limitations.

In the same context, Fig. 16 and 17 show respectively, RoCoF and output power of VSG based inverter. Again, the proposed method using FNNC is very effective in keeping the RoCoF below ± 0.2 Hz/s, while other methods face difficulties and restrictions in dealing with the uncertainty and physical constraints in the microgrid where RoCoF reaches to the large value, which may lead to relays activation.

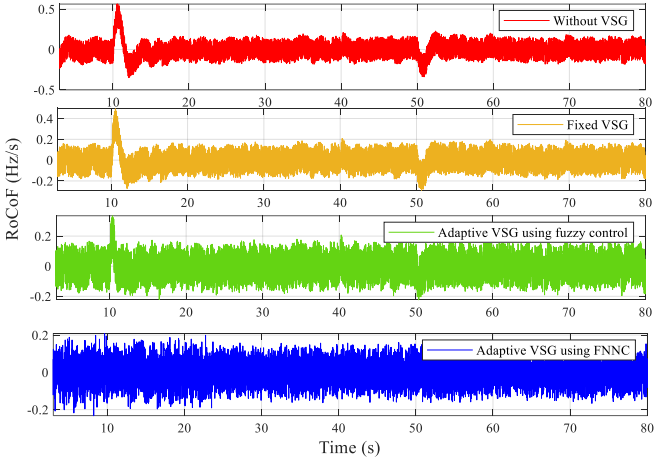
Fig. 13. RoCoF response for Scenario II (second case).

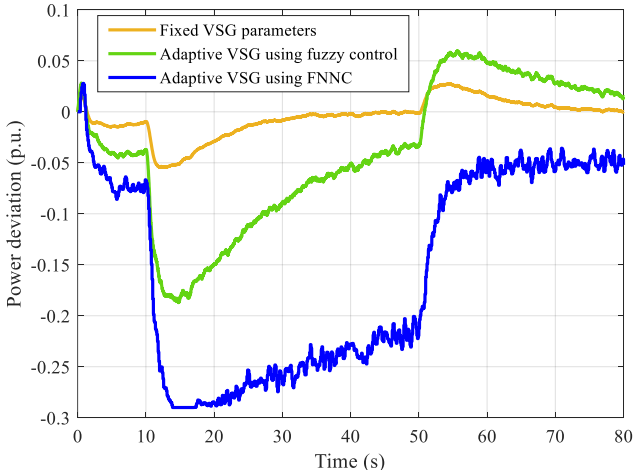
Fig. 14. Output power of VSG based inverter for Scenario II (second case).

Table V
Microgrid system parameters for Scenario III.

| Parameter | Symbol | Unit | Value |
| --- | --- | --- | --- |
| Governor time constant | $T_g$ | s | 0.15 (+50 %) |
| Turbine time constant | $T_t$ | s | 0.5 (+25 %) |
| System inertia constant | $H$ | p.u.MW s | 0.05 (– 40 %) |
| System load damping | $D$ | p.u.MW/Hz | 0.02 (+25 %) |
| Delay time | --- | s | 1.5 |
| Min. and Max. of valve gate | --- | p.u.MW | ± 0.5 |
| Speed governor dead band | --- | % | ± 0.02 |

Table VI presents the values of frequency deviation overshoot and stabilizing/recovery time of all methods used in this paper after disturbance occur at $t$ = 40 s according to Scenario I. It is clear that after the disturbance occur, our proposed method has the minimum frequency deviation as well as lesser stabilizing time compared to other methods, i.e., the highest performance and effectiveness than the rest of the methods.

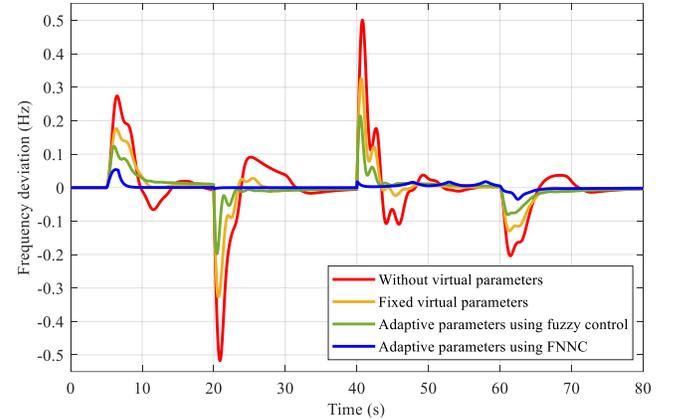
Fig. 15. The frequency deviation response under the influence of physical constraint, time delay, and uncertainties.

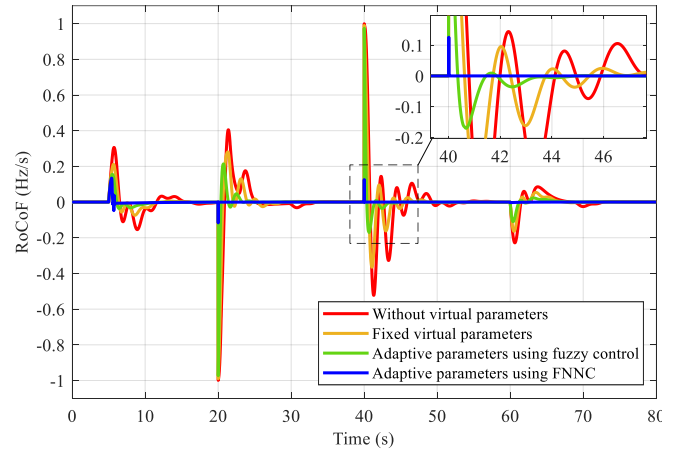
Fig. 16. RoCoF response for Scenario III.

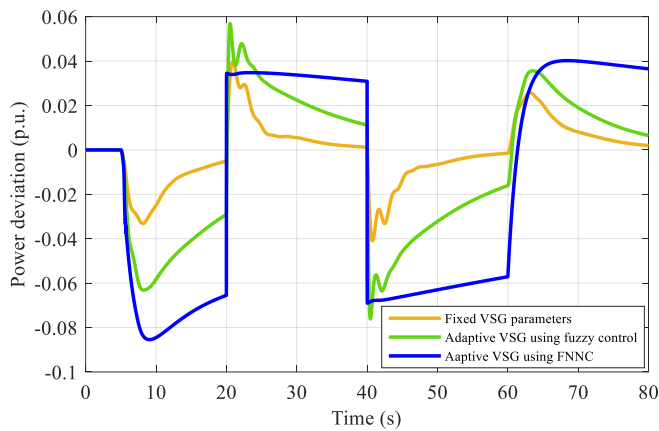

Fig. 17. Output power of VSG based inverter for Scenario III.

Table VI
Comparison of different control methods

| Methods | Frequency deviation overshoot (mHz) | Stabilizing/recovery time (s) |
|---|---|---|
| Without virtual parameters | 316 | 8 |
| Fixed virtual parameters | 230 | 10 |
| Adaptive inertia only [12] | 190 | 11 |
| Adaptive three parameters using fuzzy controller [28] | 176 | 13 |
| Proposed method (FNNC) | 71 | 2.5 |

## VI. Conclusion

This paper proposes a fuzzy neural network-based frequency control through dynamic and simultaneous adjustment the virtual parameters (inertia, damping, and droop) of a VSG-based inverter. The control unit is tasted in an isolated microgrid system based on a small-signal model to validate their effectiveness. Subsequently, we compare several aspects of the control algorithm across several scenarios. These scenarios involve external disturbances, uncertainties in system components, and variations in power from renewable energy sources.

This method offers a robust solution that enhances the dynamic frequency response of microgrid system by enabling the control unit to train itself online. The training procedure allows the controller to provide appropriate values for virtual parameters and manage the inverter to handle undesired sudden changes. As another advantage, the proposed control method does not require human experience or specific processes to build. In addition, it does not require prior knowledge of the dynamic behavior of the microgrid system. This controller can also dynamically provide optimal values for virtual parameters regardless of the disturbances or uncertainties that occur over time.

The results indicate the effectiveness of the proposed controller in significantly reducing the frequency deviation and shortening the stabilizing/recovery time. In various test scenarios, the controller maintained the frequency deviation within ±0.02 Hz. These findings confirm the high performance of the proposed method compared with traditional and fuzzy logic controller methods. Given its self-adaptive virtual dynamic control capability, which ensures independence, this method can be used to design a new generation of virtual controllers for multiple grid-forming inverters, in which no single inverter is dominant.